\documentstyle[11pt,epsfig]{article}

\renewcommand{\title}[1]{{\Large\bf\mbox{}\\\medskip#1\bigskip\medskip\\}}
\renewcommand{\author}[1]{{\large #1\smallskip\\}}
\newcommand{\address}[1]{{\em #1\medskip\\}}

\begin{document}
\begin{center}

\title{Quantum fingerprints of classical Ruelle-Pollicot resonances}
\author{ Kristi Pance, Wentao Lu and S. Sridhar$^{a}$}
\address{Department of Physics,
Northeastern University, Boston, Massachusetts 02115}


\begin{abstract}
N-disk microwave billiards, which are representative of open quantum systems, are 
studied experimentally. The transmission spectrum 
yields the quantum resonances which are consistent with semiclassical calculations. 
The spectral autocorrelation of the quantum spectrum is shown
to be determined by the classical Ruelle-Pollicot resonances, arising from the complex
eigenvalues of the Perron-Frobenius operator. 
This work establishes a fundamental connection between quantum and classical
correlations in open systems.
\end{abstract}

\end{center}



The quantum-classical correspondence for chaotic systems has been studied
extensively in the context of universality and periodic orbit contributions.
This approach has focussed on eigenvalues and eigenfunctions and their
statistical properties. Universality has been shown to arise from Random
Matrix Theory \cite{Bohigas90}, while periodic orbit contributions have been
analyzed in the semiclassical scheme for calculations of eigenvalue 
spectra \cite{Gutzwiller90} and constructions of eigenfunctions \cite{Heller}.

An entirely different approach is to consider correlations of observables.
In the classical context a probabilistic approach is best taken with
Liouvillian dynamics. In certain classical systems these have been shown to
lead to Ruelle-Pollicot (RP) resonances \cite{Ruelle86,Pollicot}, arising
from complex eigenvalues of the Perron-Frobenius operator. In open systems,
this leads to a quantitative description of the time-evolution of classical
observables, the most common being the particle density. In the quantum
context, diffusive transport has been argued to be intimately connected with
Liouvillian dynamics, not just in disordered systems where the
correspondence is made with nonlinear $\sigma -$models of supersymmetry 
\cite{Efetov} but also in individual chaotic systems which represent 
a ballistic limit.

In this paper we present a microwave experiment which demonstrates this deep
connection between quantum properties and classical diffusion. Our
experiment is a microwave realization of the well-known n-disk geometry,
which is a paradigm of an open quantum chaotic system, along with other
systems such as the Smale horseshoe and the Baker map \cite{Gaspard98}. The
classical scattering function of the chaotic n-disk system is
nondifferentiable and has a selfsimilar fractal structure. A central
property is the exponential decay of an initial distribution of classical
particles, due to the unstable periodic orbits, which form a cantor set,
hence the name {\it fractal repeller}. The experimental transmission
spectrum directly yields the frequencies and the widths of the low lying
quantum resonances of the system \cite{Lu99,Lu00}, which are in agreement
with semiclassical periodic orbit calculations \cite
{Cvitanovic93,Gaspard89,Gaspard94}. The same spectra are analyzed to obtain
the spectral wave-vector autocorrelation $C(\kappa )$ \cite{Lu99}. The wave
vector dependence of the spectral autocorrelation is shown to be completely
described by the leading RP resonances of the corresponding classical
system. The small $\kappa $ (long time) behavior of the spectral
autocorrelation provides a measure of the quantum escape rate, and is shown
to be in good agreement with the corresponding classical escape rate. For
large $\kappa $ (short time), the contribution of classical RP resonances is
observed as non-universal oscillations of the autocorrelation. Thus we are
experimentally able to observe the classical RP resonances in a quantum
experiment, for the first time.

The experiments are carried out in thin microwave structures consisting of
two highly conducting $Cu$ plates spaced $d\,\sim 6\,mm$ and about 
$55\times 55\,cm$ in area. Disks and bars also made of $Cu$ and of thickness 
$d$ are placed between the plates and in contact with them. In order to simulate 
an infinite system microwave absorber material ECCOSORB AN-77 was sandwiched
between the plates at the edges. Microwaves were coupled in and out using
terminating coaxial lines which were inserted in the vicinity of the
scatterers. All measurements were carried out using an HP8510B vector
network analyzer which measured the complex transmission ($S_{21}$) and
reflection ($S_{11}$) S-parameters of the coax + scatterer system. It is
crucial to ensure that there is no spurious background scattering due to the
finite size of the system. This was verified carefully as well as that the
effects of the coupling probes were minimal and did not affect the results.

In this essentially 2-D geometry, Maxwell's equation for the experimental
system is identical with the Schr\"{o}dinger time-independent wave equation 
$(\nabla ^{2}+k^{2})\Psi =0$ with $\Psi =E_{z}$ the $z$-component of the
microwave electric field. This correspondence is exact for all frequencies 
$f<f_{c}=c/2d=25GHz$. (Note that $k=2\pi f/c$, where $c$ is the speed of light).
It is this mapping which enables us to study the quantum properties of the
2-D systems. For all metallic objects in the 2-D space between the plates,
Dirichlet boundary conditions apply inside the metal. Similar microwave
experiments, which exploit this QM-E\&M mapping, have been used to study
quantum chaos in closed \cite{Sridhar91,microwavechaos} and open systems 
\cite{Lu99}. See \cite{Lu00} for details of the experiments.

\begin{figure}[htbp]
\center{\rule{5cm}{0.mm}}
\rule{5cm}{0.mm}
\vskip -0.9cm
\epsfig{figure=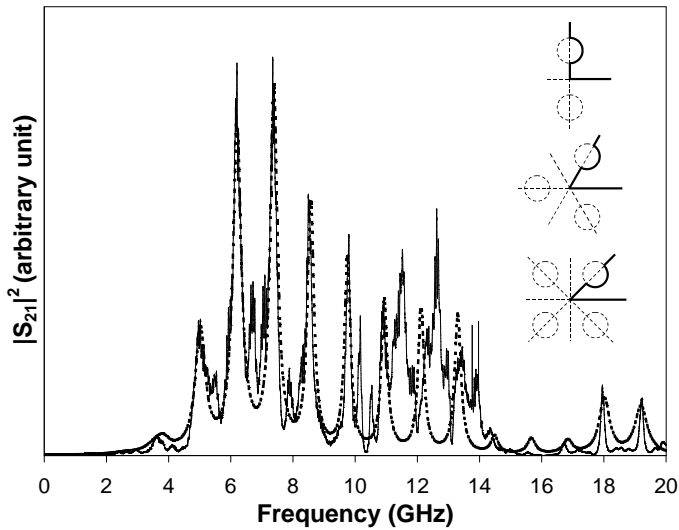,height=2.8in,angle=0}
\vskip -.3cm
\caption{Transmission spectrum $S_{21}(k)$ of the 3-disk system in the fundamental
domain with disk separation 
$R=20\sqrt{3}$ cm and radius $a=5$ cm. The dashed line is the semiclassical calculation. 
See \cite{Lu00} for details.
(Insets) Geometry of the $n$-disk open billiard ($n=2,3,4$).
The solid lines represent the fundamental domain in which
the present experiments were carried out.}
\label{3disk-tra}
\end{figure}

The transmission function $S_{21}(f)$ which we measure is the response of
the system to a delta-function excitation at point $\vec{r}_{1}$ probed at a
different point $\vec{r}_{2}$, and is determined by the wavefunction $\Psi$
at the probe locations $\vec{r}_{1}$ and $\vec{r}_{2}$. In our experiments
the coax lines act as tunneling point contacts, and hence it can be shown 
\cite{Lu00} that $S_{21}(f)=A(f)G(\vec{r}_{1},\vec{r}_{2},k)$ is just the
two-point Green's function $G(\vec{r}_{1},\vec{r}_{2},k)$. 
The scaling function $A(f)$, which represents the
impedance characteristics of the coax lines and probes, 
is sufficiently slowly varing and can 
be treated as a constant practically. Because we ensure
that the coupling to the leads is very weak, any shifts due to the leads are
negligible ($<10^{-4}$ of the resonance frequencies and widths)\cite{Lu00}.
The $n$-disk systems are investigated in the fundamental domain 
\cite{Cvitanovic93}, as shown in the inset to Fig.\ref{3disk-tra}, with angles 
$90{{}^\circ}$ ($n=2$), $60{{}^\circ}(n=3)$, and $45{{}^\circ}(n=4)$. A typical
trace for the 3-disk system is shown in Fig. \ref{3disk-tra}. See \cite{Lu00}
for details of the comparison of the resonances between experiments and 
semiclassical calculations.


\vskip 0.5cm

The spectral autocorrelation function was calculated as \cite{Lu00} 
$C(\kappa )=\left\langle |S_{21}(k-(\kappa /2))|^{2}|S_{21}(k+(\kappa
/2))|^{2}\right\rangle _{k}$. The average is carried out over a band of wave
vector centered at certain value $k_{0}$ and of width $\Delta k$. Since the
transmission function is the superposition of many resonances, 
$|S_{21}(k)|^{2}=\sum_{i}c_{i}s_{i}^{\prime }/((k-s_{i})^{2}+s_{i}^{\prime
2}) $, with $s_{i}+is_{i}^{\prime }$ the semiclassical resonances and $c_{i}$
the coupling which depend on the location of the probes, we have
\cite{Eckhardt93} 
\begin{equation}
C(\kappa )=\pi \sum_{i,j}\frac{c_{i}c_{j}(s_{i}^{\prime }+s_{j}^{\prime })}{%
(\kappa -(s_{i}-s_{j}))^{2}+(s_{i}^{\prime }+s_{j}^{\prime })^{2}}.
\label{auto-exp}
\end{equation}
In the case that there are no overlapping resonances, 
$|s_{i}-s_{j}|>>(s_{i}^{\prime }+s_{j}^{\prime })$, the small $\kappa $
behavior of the autocorrelation is $C(\kappa )\approx \pi
\sum_{i}2c_{i}^{2}s_{i}^{\prime }/(\kappa ^{2}+4s_{i}^{\prime 2})$.
According to semiclassical theory, the above sum can be replaced by a single
Lorentzian \cite{Blumel88,Jalabert90,Lewenkopf91,Lai92} 
\begin{equation}
C(\kappa )=C(0){\frac{1}{1+(\kappa /\gamma )^{2}}},  \label{lorentzian}
\end{equation}
where $\gamma =\gamma _{cl}$, the classical escape rate
with the velocity scaled to 1. The above equation was used to fit the
spectral autocorrelation for small $\kappa $ and thus obtain the value of
the experimental escape rate $\gamma _{qm}$ \cite{Lu99,Lu00}. Good agreement
of the escape rate is obtained between $\gamma _{qm}$ obtained from the
experiments and $\gamma _{cl}$ of the classical theory \cite{Lu00}.

For intermediate $\kappa $, the semiclassical prediction of Eq. (\ref
{lorentzian}) fails because of the presence of the periodic orbits, which
leads to non-universal behavior. Non-universal contributions can play in
general a crucial role in determining the overall structure of the spectral
autocorrelation, since they can be of the same order of the universal result
of Random Matrix Theory. Recently, Agam \cite{Agam00} derived a
semiclassical theory to build the connection between the nonuniversality of
the spectral autocorrelation and the classical RP resonances. 
Consider the quantum mechanical propagator \cite{Heller91}
\begin{equation}
K({\vec{r}}_1,{\vec{r}}_{2},t)=\frac{1}{2\pi\hbar i}\int G({\vec{r}}_1,
{\vec{r}}_{2},\sqrt{2m\varepsilon /\hbar^2 })
e^{-i\varepsilon t/\hbar}d\varepsilon,
\end{equation}
with $\varepsilon =\hbar^2 k^{2}/2m$. The integration is performed around 
$\varepsilon_{0}=\hbar^2 k_{0}^{2}/2m$, with 
$\Delta \varepsilon =\hbar\upsilon \Delta k$ 
and $\upsilon =\hbar k_{0}/m$ is the group velocity of the classical particle.
The integration in the $\varepsilon$ space can be changed into that in the 
$k$ space as
$K({\vec{r}}_1,{\vec{r}}_{2},t)=(\upsilon/2\pi i)e^{-i\varepsilon_{0}t/\hbar}
\int_{\Delta k}G({\vec{r}}_1,{\vec{r}}_{2},k_{0}+k)e^{-i\upsilon kt}dk$. 
The particle density is
$\rho(t)=\left|K({\vec{r}}_1,{\vec{r}}_{2},t)\right| ^{2}$.
The autocorrelation of the particle density is
$C_{\rho }(\tau )=\left\langle \rho (t)\rho (t+\tau )\right\rangle
_{t}-\left\langle \rho \right\rangle_t ^{2}$ with 
$\left\langle \rho \right\rangle_t\equiv\lim_{t\to\infty}(1/T)\int_{0}^T\rho(t)dt$.
Using the diagonal approximation, we get
$C_{\rho }(\tau)=(\Delta k\upsilon^2/4\pi^2V^2)\int d\kappa C(\kappa )e^{-i\upsilon \kappa \tau}$.
Here $V$ is the volume of the system with $V\to\infty$ for open system.
If one assumes that the above correlation is classical, one has
\cite{Ruelle78,Chris90} 
$C_{\rho }(\tau)=\sum_{i=1}^{\infty }2b_{i}e^{-\gamma_{i}\upsilon\tau}\cos
\gamma_{i}^{\prime }\upsilon\tau$, where the $c_{i}$ are the coupling coefficients,
$\gamma_{i}\pm i\gamma_{i}^{\prime }$ 
the RP resonances of the corresponding classical system in wave vector space. 
Taking the Fourier
transform of the above expression 
$\int d\tau C_{\rho }(\tau)e^{i\kappa\upsilon\tau}$, we get
\begin{equation}
C(\kappa )=\sum_{\pm ,i=1}^{\infty }\frac{b_{i}^{\prime}\gamma _{i}}
{\gamma _{i}^{2}+(\kappa \pm \gamma _{i}^{\prime })^{2}}.  \label{auto-cor}
\end{equation}
with $b_i^{\prime}=2\pi V^2b_i/\Delta k\upsilon^3$.

We now turn to the classical dynamics of the system. The classical evolution
is described by the Perron-Frobenius operator whose spectrum, known as the
RP resonances, can be calculated as the poles of the classical Ruelle 
$\zeta $-function. For the hard disk system, the classical Ruelle $\zeta $
-function is \cite{Gaspard92} 
\begin{equation}
\zeta _{\beta }(s)=\prod_{p}\,\left[ 1-\exp (sL_{p})/|\Lambda _{p}|\Lambda
_{p}^{\beta -1}\right] ^{-1},  \label{ruelle}
\end{equation}
here, $L_{p}$ the length of the periodic orbit $p$, $\Lambda _{p}$ the
eigenvalue of the monodromy matrix associated with the periodic orbits. The 
$\zeta $ -function is analytical in the half-plane ${\rm Re}\>s<-P(\beta )$,
and has poles in the other half-plane. In particular, $\zeta _{\beta }(s)$
has a simple pole at $s=-P(\beta )$. Here, $P(\beta )$ is the so-called
Ruelle topological pressure, from which all the characteristic quantities of
classical dynamics can be derived in principle. The classical escape rate is 
$\gamma _{cl}=-P(1)$. The poles of $\zeta _{\beta }(s)$ with $\beta =1$ are
calculated since they contribute to the RP resonances with the sharpest
width. 

\begin{figure}[htbp]
\center{\rule{5cm}{0.mm}}
\rule{5cm}{0.mm}
\vskip -0.9cm
\epsfig{figure=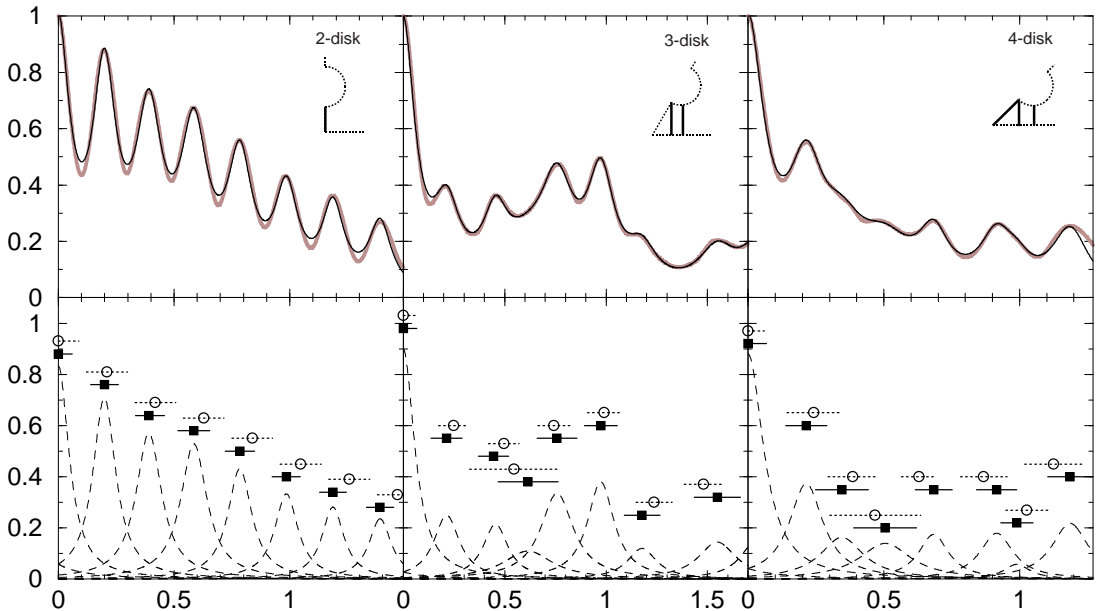,height=3.2in,angle=0}
\vskip -.2cm
\caption{Autocorrelation function $C(\kappa )$ vs $\kappa$ (cm$^{-1}$) 
and Ruelle-Pollicott resonances of the $n$-disk system in the fundamental
domain. (left). 2-disk system with $R=40$ cm and $a=5$ cm.
(center). 3-disk system with $R/a=4\sqrt{3}$ and $a=5$ cm.
(right). 4-disk system with $R/a=4\sqrt{2}$ and $a=5$ cm.
(top). The gray line is the experimental autocorrelation. The solid line
is the numerical fit to Eq. (\ref{auto-cor}). 
(bottom). The dashed lines are the decomposition of the experimental $C(\kappa )$ 
into Lorentzians. The filled squares are the position $\gamma_j$ and the bars 
the widths $\gamma_j^\prime$ of the experimental RP resonances obtained from 
the decomposition. The open circles with dotted bars are the positions 
$\gamma_j$ and width $\gamma_j^\prime$ of the predicted RP resonances 
calculated from Eq. (\ref{ruelle}). The insets in the top panels show the leading 
periodic orbits (solid lines) for the different geometries.}
\label{disk-cor}
\end{figure}

For the integrable $2$-disk system in the fundamental domain, there is just
one prime periodic orbit. We have $\zeta _{1}(s)=1-t_{0}$, where 
$t_{0}=\exp[s(R-2a)]/\Lambda $, and $\Lambda =(\sigma -1)+\sqrt{\sigma (\sigma-2)}$,
with the disk separation ratio $\sigma \equiv R/a$. The
classical scattering resonances are $s_{n}=(\ln \Lambda \pm i2n\,\pi
)/(R-2a\,)$, with $n=1,2,\cdots \,$. The classical escape rate is $\gamma
_{cl}=(\,\ln \,\Lambda \,)/(R-2a)\,.$ The semiclassical resonances in the
fundamental domain are \cite{Wirzba92} $(2n\pi +i(1/2)\ln \Lambda )/(R-2a)$
with $n=1,2,\cdots $. Substituting the semiclassical resonances into 
Eq.(\ref{auto-exp}), one may express the full two-point correlation function as 
\begin{equation}
C(\kappa )\propto \sum_{n=-\infty }^{\infty }\frac{b_{n}}
{\gamma^{2}+(\kappa +n\gamma ^{\prime })^{2}},  \label{2disk-auto}
\end{equation}
where $\gamma =\gamma _{cl}$, $\gamma ^{\prime }=2\pi /(R-2a)$. On the other
hand, since the RP resonances of the system are 
$(\ln \Lambda +i2n\pi)/(R-2a)$, if one puts these resonances into 
Eq. (\ref{auto-cor}), the above
expression (\ref{2disk-auto}) follows immediately. The experimental RP
resonances are obtained from the autocorrelation by fitting it with 
Eq. (\ref{2disk-auto}). Since the transmission coefficient $S_{21}(f)$ and also the
couplings $c_{i}$ of the quantum resonances depend on the location of the
two probes, so does the coupling $b_{n}$ of the classical RP resonances. The
comparison between experiment and theory is shown as Fig. \ref{disk-cor} (left).

\vskip 0.4cm

For the chaotic $n$-disk system, making use of the cycle expansion \cite
{Cvitanovic89} and also the symmetry factorization of the classical Ruelle 
$\zeta $-function, the RP resonances can be calculated very accurately \cite
{Gaspard92}. For the $3$-disk system in the fundamental domain, the
classical RP resonances are calculated by using 8 prime periodic orbits up
to period 4. The calculation is very accurate for ${\rm Re}\>s<0.8$ 
\cite{Gaspard92}. The classical RP resonances can also be obtained experimentally
by fitting the autocorrelation with Eq. (\ref{auto-cor}). Because of the
finite range of the experiment spectrum, only the first 8 or 9 RP resonances
with small real part were obtained. The experimental autocorrelation with
the theoretical prediction are shown as Fig. \ref{disk-cor} (center). 

For the $4$-disk system in the fundamental domain, the classical RP
resonances are calculated by using 14 prime periodic orbits up to period 3.
The experimental autocorrelation with the theoretical prediction are shown
as Fig. \ref{disk-cor} (right). 

The agreement between the experimental RP resonances and the theoretical
ones for the 2-disk sytem is 6\% for the positions $\gamma _{j}$
and better than 30\% for the widths $\gamma _{j}^{\prime }$,
is 7\% for $\gamma _{j}$ and 11\% for $\gamma_{j}^{\prime }$ 
for the 3-disk sytem, and is 8\% for $\gamma _{j}$ 
and 17\% for $\gamma _{j}$ for the 4-disk sytem.
We note that these agreements, in particular the wave-vector locations 
$\gamma _{j}$ should be considered as very good. The principal sources for
the residual discrepancies are the nonideality of the absorbers, small
symmetry-breaking perturbations and the suppression of some resonances at
the neighborhood where the antennas are coupled which affects the
autocorrelation function, therefore the position and the widths of RP
resonances. Also very broad resonances are difficult to identify and can
lead to an apparent enhancement of the observed widths, which can possibly
account for the systematically larger widths that are observed.

Our investigation clearly demonstrates that the whole spectral
autocorrelation can be understood completely in terms of the classical RP
resonances. The meaning of these RP resonances in the classical context can
be understood as follows. If one shoots particles toward the hard disk
scatterer, the number of particles that will remain in the scattering region
will decay as $N(t)=\sum_{i}a_{i}e^{-\alpha_{i}t}.$ Besides the general
exponential decay at $\alpha_{0}=\upsilon\gamma _{cl}$, 
there are oscillations due to the fact that the RP resonances 
$\alpha_{i}=\upsilon(\gamma_i\pm i\gamma^{\prime}_i)$ are not always real \cite
{Gaspard92} as contrasted with the purely diffusive system. Taking the
Fourier transform of $N(t)$, one can identify the Lorentzians in the
spectrum with the RP resonances. Our work demonstrates that suitable quantum
correlations diffuse just like classical observables in an open system.

It is remarkable that the same experiment yields both the quantum resonances
and the classical RP resonances. Thus we have demonstrated experimentally
the profound connection between quantum properties and classical diffusion.
This connection is best seen in open quantum systems. While we have studied 
the model n-disk geometry, the results have broad implications for arbitrary chaotic
geometries. The results of this work also have wider implications in a
variety of phenomena in different fields in physics, such as
photodissociation of atoms \cite{Agam99}, nuclear decay \cite{Bauer90},
electronic transport, fluid dynamics, and acoustic and electromagnetic propagation.

We thank P. Pradhan for useful discussions. This work was supported by
NSF-PHY-9722681.

$^{a}$ electronic address : srinivas@neu.edu.




\end{document}